\documentstyle[12pt]{article}
\def\ymin{y_{\rm min}}
\begin{document}
\begin{titlepage}
\vspace*{-1cm}
\begin{flushright}
DTP/95/38   \\
UR-1420 \\
ER-40685-869 \\
May 1995 \\
\end{flushright}
\vskip 1.cm
\begin{center}
{\Large\bf
On Gluon Radiation in ${\rm t\bar{t}}$ Production
and Decay}
\vskip 1.cm
{\large Lynne H. Orr}
\vskip .2cm
{\it Department of Physics, University of Rochester \\
Rochester, NY 14627-0171, USA }\\
\vskip   .4cm
{\large  T. Stelzer}
\vskip .2cm
{\it Department of Physics, University of Durham \\
Durham DH1 3LE, England }\\
\vskip   .4cm
and
\vskip .4cm
{\large  W.J. Stirling}
\vskip .2cm
{\it Departments of Physics and Mathematical Sciences, University of Durham \\
Durham DH1 3LE, England }\\
\vskip 1cm
\end{center}
\begin{abstract}
Understanding the pattern of gluon radiation in $t \bar t$ production
processes is important for making an accurate determination of the top
mass from the measurement of its decay products. In a recent paper we
showed that the exact matrix element and parton shower (HERWIG)
calculations gave very different results for the distribution of gluon
jets in $t \bar t$ production at the Tevatron $p \bar p$ collider.  By
repeating the calculation for the simpler $e^+e^- \to b \bar b W^+W^-g$
process, we reveal even more dramatic differences between the two
approaches.  We conclude that there are significant differences in
gluon radiation between HERWIG and the matrix element calculation in
regions of phase space where one would expect agreement.
\end{abstract}
\vfill
\end{titlepage}
\newpage

Having established the existence of the top quark through its
production in $p \bar p$ collisions at the Fermilab Tevatron
\cite{CDFTOP,D0TOP}, the CDF and D0 collaborations must now accurately
measure the top mass.  To do so requires understanding gluon radiation
in top events.  Because top is observed only through its decays,
measuring its mass requires reconstructing the momenta of its decay
products and comparing measured invariant mass distributions with
those predicted by theory. And because gluons can be radiated in top
production {\it and} decay, top events often contain extra jets which
may or may not need to be counted among the decay products.  While
this cannot be decided on an event-by-event basis, in principle the
problem of extra jets can be dealt with as long as their distributions
are properly understood.

In previous work \cite{OS,OSS} we studied the distributions of extra
jets in $t \bar t$ production and decay at the Tevatron $p \bar p$
collider.  In \cite{OSS} we performed a complete tree-level
calculation of the process $p \bar p \to b W^+ \bar b W^- + $ jet $+
X$, in which we included all contributions from gluons emitted in the
production and decay stages as well as all top width effects and spin
correlations.  We compared our exact matrix element results to those
obtained using the parton-shower Monte Carlo program HERWIG
\cite{HERWIG}, which is widely used in the experimental analyses.  We
found a significant discrepancy between the two calculations.
Although we were unable to identify the reason for the discrepancy,
the differences in jet distributions seemed to indicate a relative
lack of gluons radiated in top decay compared to top production in
HERWIG.

In this paper we pursue further the comparison of matrix element (ME)
and parton shower (PS) calculations of gluon radiation in top
production and decay.  In light of the subtle issues associated with
making ME--PS comparisons, and of the complications associated with
hadronic top production, it is necessary to investigate the
differences in a systematic way.  Therefore we focus here on the
simpler case of $e^+e^-$ collisions, where many of the complications
of $p \bar p$ collisions, such as initial state radiation, are absent,
and where a cleaner comparison can be performed.  Then, armed with our
$e^+e^-$ results, we return to a discussion of the hadronic case.

Comparing a fixed order calculation with a full parton shower program
presents several challenges.  Most notable is that the ME calculation
produces only partons directly from the hard process, whereas in the
PS calculation, any number of partons can be present as a result of
showers, and multiple gluon effects can be important.  In order to
develop a meaningful method for comparison, we begin by studying the
well-understood process $e^+e^-\to q\bar q g$, where $q$ is a light
quark. Next, we make the same comparison but including a non-zero
quark mass ($e^+e^-\to Q\bar Q g$). From these processes a useful
method of comparison is found, and the regions of applicability for
the different calculations are studied.  Having understood the simpler
processes, we then consider $e^+e^-\to b W^+ \bar b W^- + g$ via top
pair production.  Cuts are kept to the minimum necessary to avoid
singularities.

In each case, the final state hadrons in the ME calculation are two
quarks and a single extra gluon (we ignore the $W$'s from top decay;
they can be considered to decay leptonically).  In the PS calculation
the final state contains two quarks plus any number of showered
partons.  In order to make the ME--PS comparison, we must use some
algorithm to combine showered partons in the PS calculation so that we
are left with at most three jets.\footnote{The PS results were
obtained using HERWIG \cite{HERWIG} (v5.8), with hadronization and $b$
quark decays turned off.}  We use the Durham successive combination
algorithm \cite{DURHAM}, which has the advantages \cite{DURHAM2} that
it reproduces LEP data well and lends itself to direct comparison with
ME calculations.

We apply the Durham algorithm as follows.  For each
pair of partons (or jets) $i$ and $j$, we compute the quantity
\begin{equation}
y_{ij} \equiv {{2\min(E_i^2,E_j^2)(1-\cos\theta_{ij})}
              \over{s}},
\label{y}
\end{equation}
where $E_{i(j)}$ is the energy of parton $i (j)$, $\theta_{ij}$ is the
angle between them, and $s$ is the process center-of-mass energy.  The
pair with the smallest $y_{ij}$ are combined into a single jet with
four-momentum $p=p_i+p_j$.  The jet replaces the partons in the
$y_{ij}$ computations and we repeat the process until no more than
three jets remain.  Note that for small angles, the numerator of
$y_{ij}$ is simply the $k_T^2$ of the less energetic parton with
respect to the more energetic one.  Hence the algorithm causes the
partons with the smallest relative transverse momentum to be combined.
In addition, the dimensionless variable $y$ is somewhat like an
angular variable, so that this method is analogous to the angular cone
algorithm used in our comparisons for $p \bar p$ \cite{OSS}.

After using the jet algorithm, we have three-`jet' events from both PS
and ME calculations, and they can now be compared.  To do so, we make
further use of the Durham variable: for each event, we plot the
distribution as a function of the smallest value of $y$ ($\equiv
\ymin$) for all pairings of the remaining three jets.  Because large
values of $\ymin$ correspond to large relative $k_T$, {\it i.e.},
large angles and energies, the ME and PS results should agree well in
this region.  Small values of $\ymin$ bring us into the soft/collinear
regime, where the ME result diverges as $\ymin \to 0$.  In this region
multiple gluon effects become important, and we expect the PS result,
which takes these into account via resummation, to remain finite.

This behavior is illustrated in Fig.~{1(a)}, which shows the $\ymin$
distribution for $e^+e^-\to q\bar q g$ (massless quarks) with a
center-of-mass energy of 100~GeV.  Since we are primarily interested
in shapes of distributions, the normalization of the ME calculation is
chosen (via the $\alpha_s$ scale) so that the distributions for
$e^+e^-\to q\bar q g$ agree at large values of $\ymin$.  We see, then,
that ME and PS agree quite well for large $\ymin$, not only in
normalization, which we have fixed, but in shape, which we are not
free to manipulate.  As $\ymin$ decreases, the ME distribution rises,
diverging as $\alpha_s \ln^2(1/\ymin)$, but we see the multigluon
effects begin to show in the PS curve, which turns over and remains
finite.

The effect of giving the final quark a mass is shown in Fig.~{1(b)}
for the process $e^+e^-\to b\bar b g$.  Note that the $b$ mass has
weakened the ME divergence, and the distribution for all $\ymin$ is
less steep than in the massless case.  The PS calculation again
reproduces the ME result (including the rounding due to the mass)
quite well in the large $\ymin$ region, and we see the two results
begin to disagree as multigluon effects come in for smaller $\ymin$.

Satisfied that we understand the above simple cases, we turn to top
production.  The ME and PS calculations performed here for $e^+e^-$
production of top are similar to those performed in \cite{OSS}.  We
take $m_t=175$ GeV.  To obtain the ME results we have performed an
exact parton-level calculation of $e^+e^-\to b W^+ \bar b W^- + g$,
including top width effects, using helicity amplitudes generated by
the MadGraph \cite{MADGRAPH} package.  The PS results were obtained
using HERWIG \cite{HERWIG} (v5.8), with hadronization and $b$ quark
decays switched off.  As in the case of hadronic top production, we do
not consider radiation off the $W$ decay products, and the $W$'s can
be considered to decay leptonically.

We proceed with the jet algorithm as above, and compare the ME and PS
$\ymin$ distributions for center-of-mass energy 420 GeV in
Fig.~{2}.\footnote{This energy is chosen to approximate the average
subprocess center-of-mass energy for $t \bar t$ production in $p \bar
p$ collisions at the Tevatron.}  The difference is quite dramatic.
The ME curve displays further rounding as one might expect from a
higher mass quark or from the lack of correlation between the final
state $b$ quarks and gluons radiated from the $t$ or $\bar t$.  The
behavior of the PS curve is unexpected.  We see the usual turn-over at
small values of $\ymin$, but the large $\ymin$ behavior is surprising.
Not only does PS not reproduce ME; its shape is so different that
there is no shift in normalization that would cause the two to agree.

Given that the matrix element calculation is of fixed order in
perturbation theory, one expects that its region of validity would
start at large values of $\ymin$ and extend to the left to the region
where we expect multiple gluon emissions to become important.  Since
the parton shower has these multiple emissions, but uses an angular
ordering algorithm, we expect its region of validity to be from small
values of $\ymin$ and extend to the right.  The fact that there is
almost no region of overlap is very disturbing.

We have been unable to pin down the reason for the above discrepancy.
Although it is in principle difficult to compare fixed order (ME) with
approximate all-orders (PS) calculations, the fact that the comparison
works well for {\it light} quarks suggests that this is not the
explanation for the difference. Furthermore, there is no reason why a
matrix element calculation which works well for light quarks should
fail for heavy quarks, provided of course that appropriate cuts are
used to avoid the soft and collinear regions. Following the results of
Ref.~\cite{OSS}, it has been suggested \cite{FRIXIONE} that the
observed differences between the PS and ME calculations were due to a
choice of cuts in the latter that were too sensitive to the infra-red
region, thus implying that important higher-order multi-gluon
contributions were missing from the ME calculation. However (see
below) increasing the jet $E_T^{\rm min}$ cut in the calculation of
Ref.~\cite{OSS}, and thus moving to a phase-space region where the ME
approach should work even better, does not resolve the problem.

We are therefore forced to conclude that the implementation of very
heavy quark production and/or decay in HERWIG may not be correct. One
can perhaps shed further light on this by comparing the two
calculations for {\it stable} top production, thus testing the
`production' part of the jet cross section in each case.  The result
of such a comparison is shown in Fig.~3, where the distribution in
$\ymin$ is shown for stable $m_t = 175$~GeV $t \bar t g$ production in
$e^+e^-$ collisions at $\sqrt{s} = 420$~GeV.  In contrast to Fig.~1
(essentially the same distribution but for light quarks), there is a
marked difference between the two calculations in the medium-small
$\ymin$ region.  Note that in this case there is no collinear
singularity and so the divergence as $\ymin \to 0$ is weaker ($\sim
\alpha_s \ln\ymin$) in the ME distribution. It is difficult to see how
higher-order multi-gluon contributions could produce such a pronounced
{\it increase} at $\ymin \sim 10^{-4}$ in the PS calculation.  By
comparing Figs.~2 and 3, we see that the difference in the
`production' part can also be seen in the full `production $+$ decay'
distribution.

Finally, we return to the case of hadronic top production.  We recall
that the discrepancy we found in \cite{OSS} is similar to that
observed in $e^+e^-$ collisions (Fig.~2).  Essentially it amounts to
a relative deficit of decay stage radiation in the PS calculation.
This effect is both enhanced and clarified in the study in $e^+e^-$
because of the lack of initial state radiation, and the cleaner
environment.  As a final confirmation of this fact, we replot in
Fig.~4 the distributions in the phase-space separation variable
$\Delta R_{bj}$ in $p \bar p$ collisions at $\sqrt s = 1.8$~TeV from
the ME and PS (HERWIG) calculations, with a very large cut on the
transverse energy of the jets. In this region, the tree-level ME
calculation should certainly be a good approximation \cite{FRIXIONE}.
Again we observe a large discrepancy between HERWIG and the matrix
element calculation.  It is difficult to estimate the effect of this
difference on top quark analyses, and in particular on the
measurement of the top mass, but clearly every effort should be made
to ensure that the event simulators used to study the top quark do
indeed constitute a good approximation to the underlying physics.

\vskip 1truecm

\noindent Two of us (TS,WJS) are grateful to the UK PPARC
for financial support.  Useful discussions with Nigel Glover,
Michelangelo Mangano, Stephen Parke and Bryan Webber are acknowledged.
This work was supported in part by the U.S.\ Department of Energy,
under grant DE-FG02-91ER40685 and by the EU Programme ``Human Capital
and Mobility'', Network ``Physics at High Energy Colliders'', contract
CHRX-CT93-0537 (DG 12 COMA).
\goodbreak

\goodbreak

\section*{Figure Captions}
\begin{itemize}

\item [{[1]}] (a)~ $e^+e^- \to q \bar q g$ at $\sqrt s = 100$~GeV:
  PS (HERWIG) versus ME as a  function of $\ymin$.
A cut of $\ymin > 2 \times 10^{-3}$ has been placed on the matrix element
calculation. \\
(b)~ $e^+e^- \to b \bar b g $ at $\sqrt s = 100$~GeV:
 PS (HERWIG) versus ME as a function of $\ymin$.
A cut of $\ymin > 2 \times 10^{-4}$ has been placed on the matrix element
calculation.
\item [{[2]}] $e^+e^- \to b W^+ \bar b W^-g$ at $\sqrt s = 420$~GeV
with $m_t = 175$~GeV:
PS (HERWIG) versus ME as a function of $\ymin$.
\item [{[3]}] $e^+e^- \to t \bar t g$ at $\sqrt s = 420$~GeV
with $m_t = 175$~GeV and no top decay: PS (HERWIG) versus ME
as a function of $\ymin$.
\item [{[4]}]  $p \bar p  \to b W^+ \bar b W^-g + X$ at $\sqrt s =
1.8$~TeV with $m_t = 175$~GeV: PS (HERWIG) versus ME,
distribution in $\Delta R_{bj}$  for $E_T^{\rm min} > 20~{\rm GeV}$,
$\Delta R_{bj} > 0.4$.

\end{itemize}

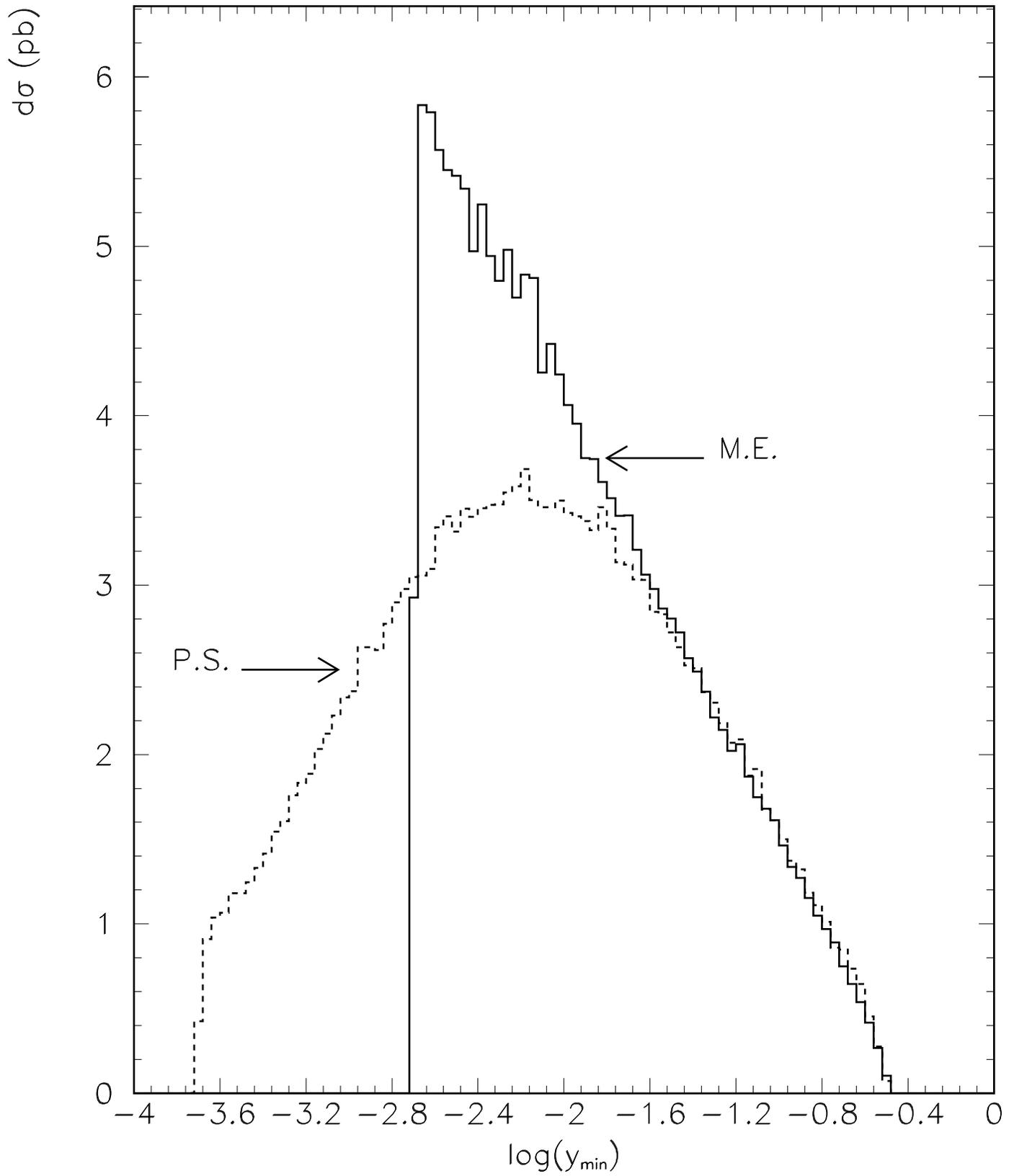
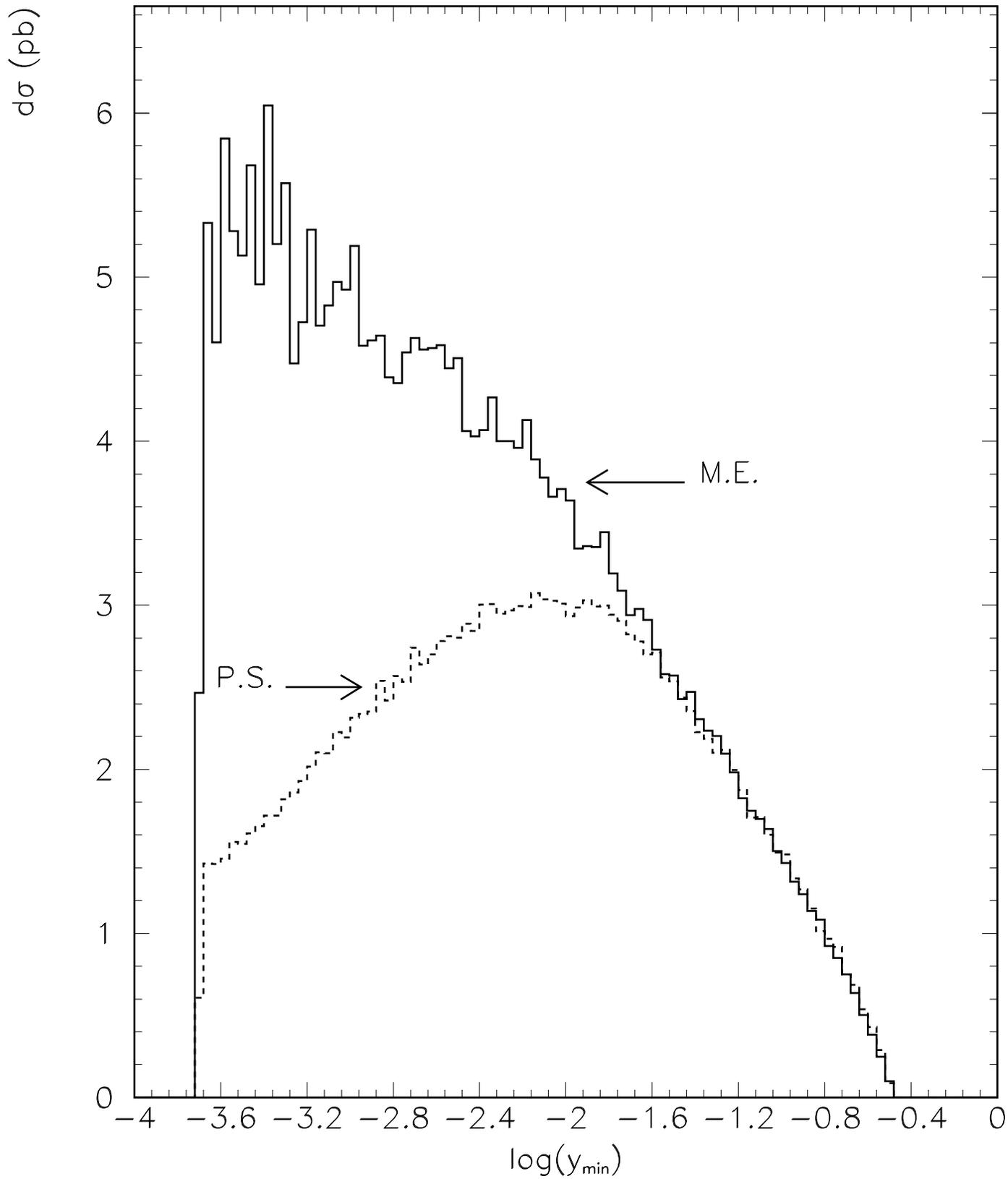
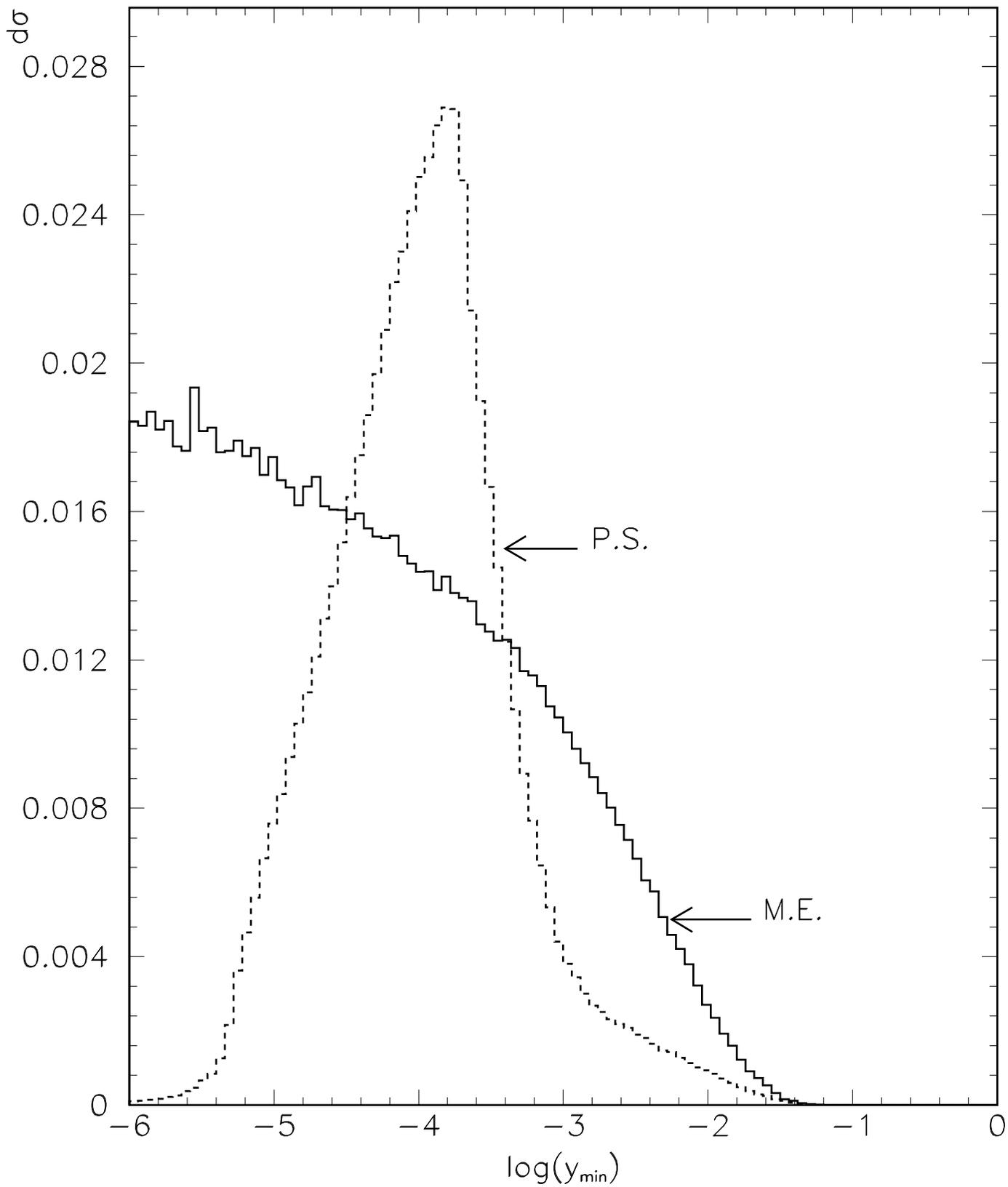
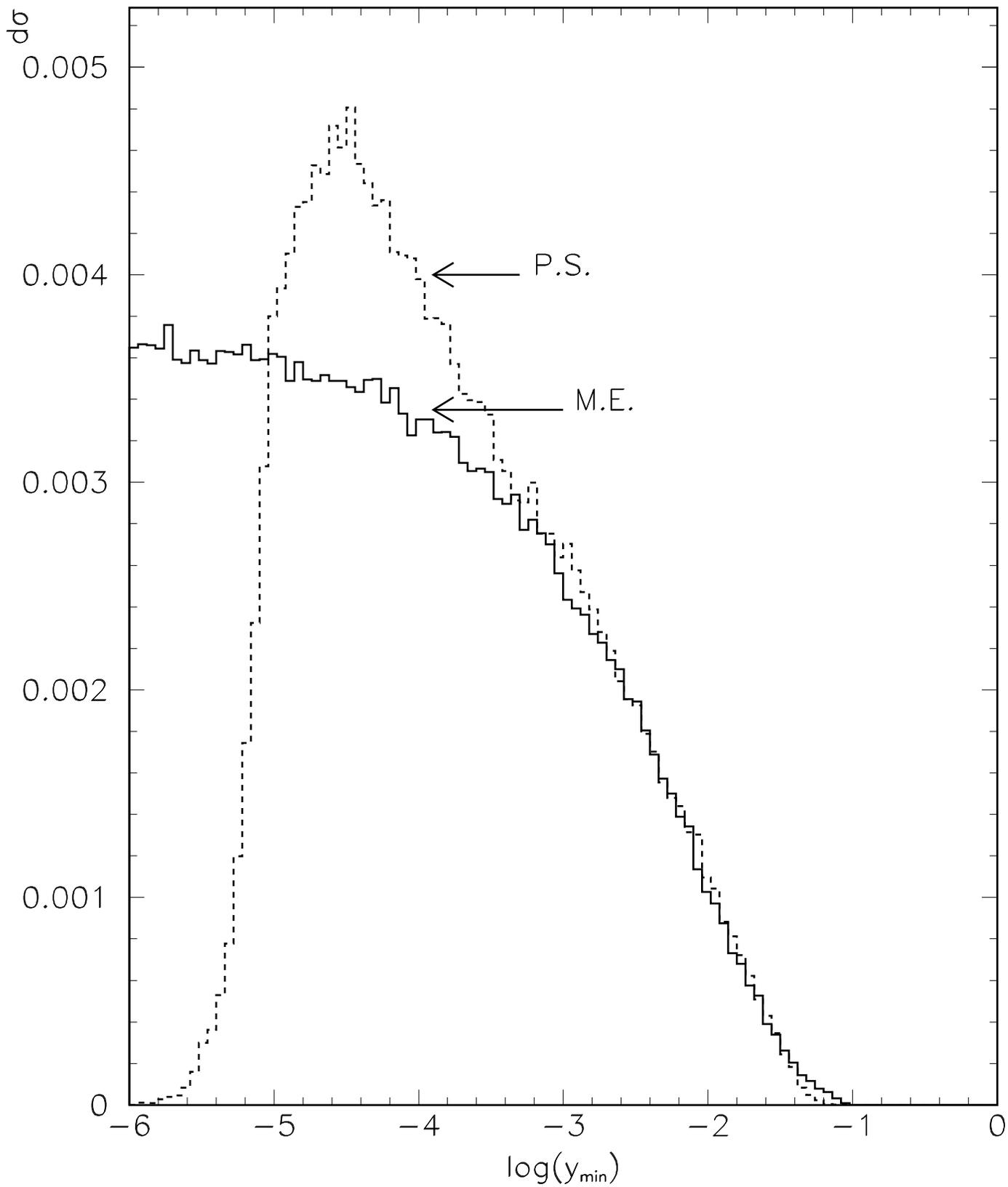
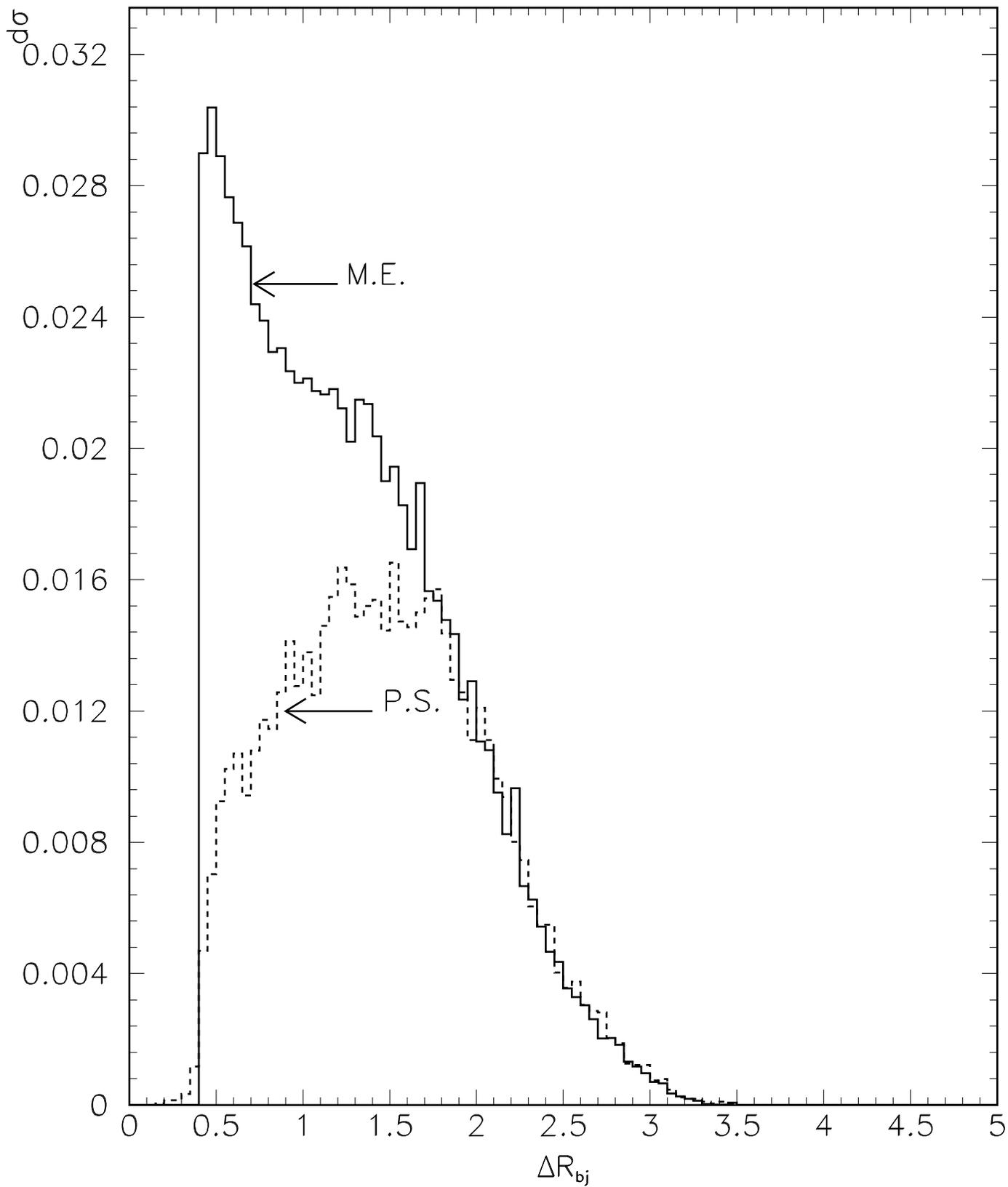
\end{document}